\documentclass[prc,twocolumn,nofootinbib,showpacs,floatfix,letterpaper]{revtex4}

\usepackage{latexsym}
\usepackage{pst-plot,pstcol,multido,graphicx}
\usepackage{amsmath,amssymb}
\usepackage{ulem}
\usepackage{multirow}
\usepackage[body={7.33in,9.45in},top=0.75in]{geometry}
\graphicspath{{./figures/}}
\newcommand{\VPO}[1]{V_{12}}
\begin{document}


\title{Global microscopic calculations of ground-state spin and parity 
for odd-mass nuclei}

\author{L. Bonneau\textsuperscript{1}, P. Quentin\textsuperscript{1,2} and 
P. M{\"o}ller\textsuperscript{1}}

\affiliation{\textsuperscript{1}Theoretical Division, Los Alamos
National Laboratory, Los Alamos, NM 87545, USA \\ 
\textsuperscript{2}Centre d'Etudes Nucl\'eaires de Bordeaux-Gradignan, 
Universit\'e Bordeaux 1 -- IN2P3/CNRS, BP 120\\
33175 Gradignan cedex, France
}

\date{\today}

\begin{abstract}

Systematic calculations of ground-state spin and parity of odd-mass nuclei 
have been performed within the Hartree--Fock--BCS (HFBCS) approach 
and the Finite-Range Droplet Model for nuclei for which experimental 
data are available. The unpaired
nucleon has been treated perturbatively, and axial and left-right
reflection symmetries have been assumed. As for the HFBCS approach, 
three different Skyrme forces have been used in the particle-hole 
channel, whereas the particle-particle matrix elements have been
approximated by a seniority force. The calculations have been done 
for the 621 nuclei for which the Nubase 2003 data set give assignments 
of spin and parity with strong arguments. The agreement of both spin
and parity in the self-consistent model reaches about 80\% for
spherical nuclei, and about 40\% for well-deformed nuclei regardless 
of the Skyrme force used. As for the macroscopic-microscopic approach, 
the agreement for spherical nuclei is about 90\% and about 40\% for 
well-deformed nuclei, with different sets of spherical and deformed nuclei 
found in each model.

\end{abstract}

\pacs{
{21.10.Hw},
{21.60.Jz}
}

\maketitle


To describe the rich variety of elementary modes of nuclear 
excitations and build up some confidence in the approaches used 
for that purpose, it is important to assess in a quantitative systematic 
way the quality of the reproduction of experimental 
static properties. We choose here to compare the results of 
Skyrme--Hartree--Fock--BCS (hereafter referred to as HFBCS) 
calculations of ground-state (GS) spin and parity with experimental 
values assigned with strong arguments available in the Nubase 
2003 database~\cite{Nubase2003} for 621 odd-Z and odd-N nuclei. 
We undertake such a global study with three different effective 
interactions, namely the SIII~\cite{SIII}, SkM*~\cite{SkMstar} and 
SLy4~\cite{SLy4} parameterizations of the Skyrme effective 
nucleon-nucleon interaction in the mean-field channel, together 
with a seniority force in the pairing channel. The results are also 
compared with those obtained previously in the Finite-Range Droplet Model 
(FRDM)~\cite{Moller_spins1990}

We would like to emphasize the global character of the present work. 
Even though some local studies (in limited mass regions) of 
the GS spectroscopic properties of odd-mass nuclei 
using such phenomenological interactions have been carried out over 
the years (see, e.g., Refs.~\cite{Daniere,Libert_Meyer_Quentin}), 
we want to check as completely as possible the relevance 
of the obtained results by limiting ourselves to two observables, the 
spin and the parity of the ground state. The present paper is a brief 
report on these first results. More details on the approximations and 
technical methods will be given in a forthcoming study of exotic 
neutron-rich nuclei. 

Our self-consistent mean-field approach including pairing correlations 
relies on the Hartree--Fock and BCS approximations. They have been 
implemented in earlier fission studies for even-even 
heavy~\cite{papier_fission_EPJA,papier_Fm} and light 
nuclei~\cite{papier_fission_PRC} with the Skyrme interaction 
in its SkM* parameterization which successfully described 
potential-energy surfaces in terms of constraints on deformation. 
As for the pairing residual interaction, we use the seniority force in 
the $T = 1$ channel, neglecting the neutron-proton correlations given 
the large value of $T_z =(N-Z)/2$, and we calculate the strength 
with a given pairing window in the same approach as the one used by 
M\"oller and Nix~\cite{Moller_Nix_pairing}. In the present 
study we choose to include in the pairing window all the 
single-particle states below the Fermi level and those lying 
$\Delta\epsilon=6$~MeV at most above the Fermi level. As in 
Ref.~\cite{Bonche_NPA_1985} we have included a smooth cut-off as a 
function of the single-particle state energy with a diffuseness 
parameter $\mu=0.2$~MeV.

We assume that GS nuclear shapes possess left-right reflection 
and axial symmetries. Global studies \cite{Moller_mass-table,Moller_triax} 
indicate that close to the valley of $\beta$-stability this is true 
except for a few small regions. Therefore the projection $K$ of the total 
angular momentum of the nucleus on the $z$-axis, chosen to coincide with 
the intrinsic symmetry axis, and the parity $\pi$ are good quantum numbers. 
In mean-field models this also holds for the single-particle states 
since the mean-field possesses these symmetries. For odd-mass nuclei, 
the time-reversal symmetry is broken but axial symmetry is assumed to 
be preserved. Depending on the intrinsic nuclear deformation, different 
assumptions for the coupling schemes have to be considered. 

\begin{table}
\caption{\label{tab_res}
Agreement of GS spin and parity calculated with three different 
Skyrme interaction in the HFBCS approach and with the FRDM model with 
respect to Nubase 2003 data~\cite{Nubase2003}, for spherical (``Sph.''), 
well-deformed nuclei (``Def.'') and all of them (``Total''). For each 
Skyrme interaction, we indicate in parenthesis on the first line the 
percentage of agreement obtained when considering either of the two 
lowest quasiparticles and on the second line the number of nuclei 
for which the GS spin and parity of the second lowest quasiparticle 
agrees with the experimental data.
}
\begin{center}
\begin{tabular}{cccc}
\hline\hline
Model & Sph. & Def. & Total  \\
\hline
SIII & 83.9\% (90.8\%) & 40.5\% (61.5\%) & 66.4\% (79.0\%) \\
     & 183(+15)/218 & 60(+31)/148 & 243(+46)/366 \\
SkM* & 76.2\% (89.2\%) & 37.5\% (61.8\%) & 63.3\% (80.0\%) \\
     & 218(+37)/286 & 54(+35)/144 & 272(+72)/430 \\
SLy4 & 77.8\% (85.8\%) & 39.3\% (60.7\%) & 64.1\% (77.6\%) \\
     & 186(+19)/239 & 57(+32)/140 & 243(+51)/379\\
FRDM & 90.9\% & 43.1\% & 54.4\% \\
     & 90/99  & 137/318  & 227/417\\
\hline\hline
\end{tabular}
\end{center}
\end{table}
\begin{figure*}
\begin{center}
\vspace*{1cm}
\hspace*{-5cm}
\rotatebox{-90}{\includegraphics[width=0.65\textwidth]{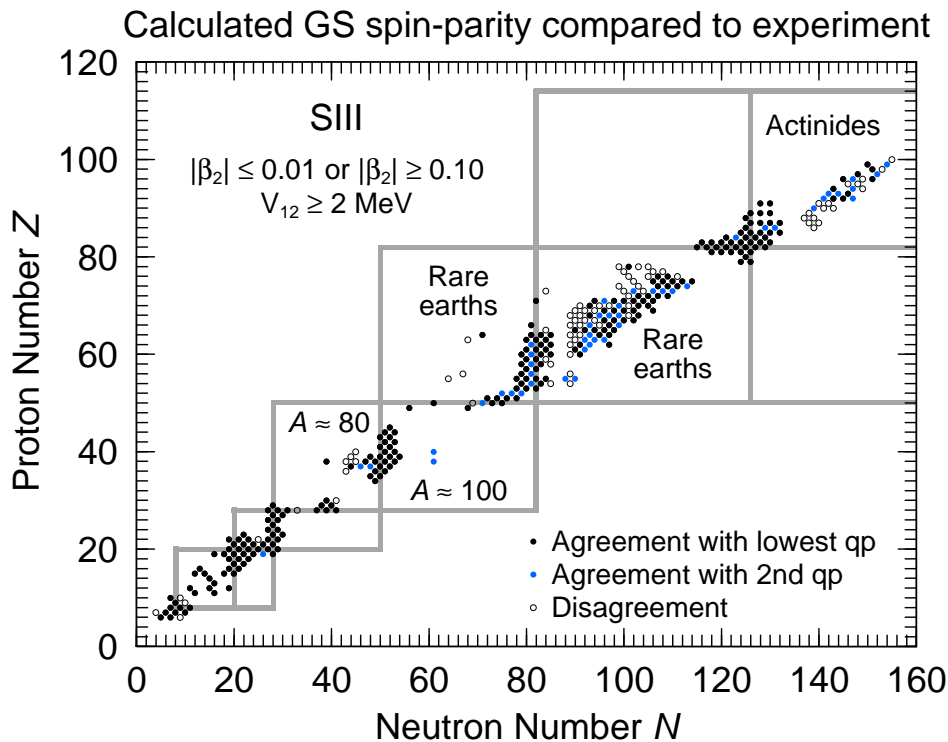}}
\end{center}
\vspace*{-3.5cm}
\caption{(color online). Comparison with experimental data of the calculated 
GS spin and parity of odd-mass nuclei for which $\beta_2 \protect\notin
[0.01;0.1]$ and $V_{12}\geqslant 2$~MeV throughout the nuclide chart
within the HFBCS model using the SIII interaction.
\label{map_agr_SIII_ALL_2}}
\end{figure*}
The nuclei for which the lowest-energy configuration is deformed are 
assumed to be rigid rotors. We can therefore describe the coupling between 
the unpaired particle and the rotation of the even-even core in the 
rotor-plus-quasiparticle approximation using HFBCS single-particle states 
as in Ref.~\cite{Libert_Meyer_Quentin}, and we assume that vibrations 
are ``frozen'' in the deformed well (zero phonon). To determine the 
lowest-energy quasiparticle in the HFBCS approach, we use the equal-filling 
approximation (see Ref.~\cite{Perez_fission_odd} and references quoted 
therein) to obtain a time-even state having the desired odd number of 
particles on average and then we create the lowest-energy quasiparticle 
on this state. 

For a spherical nucleus, the GS spin $J$ and parity $\pi$ are assumed to 
be those of the single nucleon, deduced from the quantum numbers 
($n$, $\ell$, $j$, $m$ with usual notation) of the last filled orbit, 
namely $J=j$ and $\pi=(-1)^{\ell}$. In our mean-field approach, this nucleon 
occupies the first empty level of the self-consistent one-body potential 
above the lowest-energy levels occupied by the nucleons of the even-even 
core. 

In the FRDM model the single-particle states are obtained by 
diagonalization of the folded-Yukawa one-body Hamiltonian associated 
with the GS shape as explained in detail in Ref.~\cite{Bolsterli}. 
The ground state of the nucleus is approximated by the Slater 
determinant built from the lowest-energy single-particle states 
(exhibiting the Kramer's degeneracy). The spin and parity of an odd 
nucleus are therefore those of the level occupied by the unpaired 
nucleon, that is, the highest occupied level for the particle type 
in odd number~\cite{Moller_Nix_Kratz}. This approximation is 
implemented for spherical as well as deformed nuclei. 
\begin{figure*}[t]
\begin{center}
\vspace*{1cm}
\hspace*{-5cm}
\rotatebox{-90}{\includegraphics[width=0.65\textwidth]{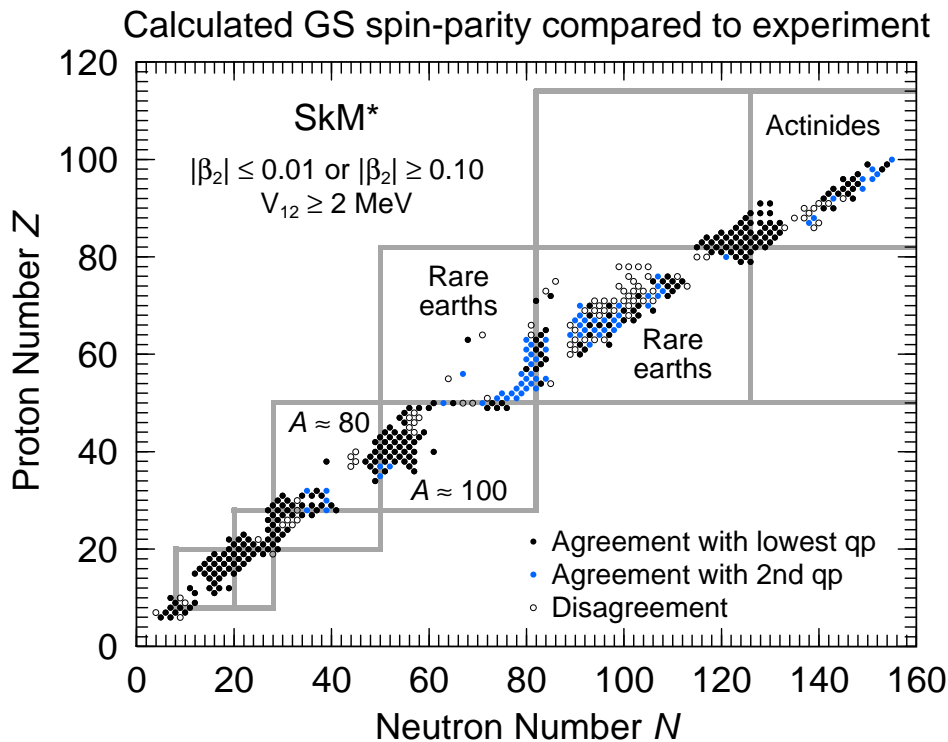}}
\end{center}
\vspace*{-3.5cm}
\caption{(color online). Same as Fig.~\ref{map_agr_SIII_ALL_2} with the 
SkM* interaction.\label{map_agr_SkM*_ALL_2}}
\end{figure*}
\begin{figure*}[b]
\begin{center}
\vspace*{1cm}
\hspace*{-5cm}
\rotatebox{-90}{\includegraphics[width=0.65\textwidth]{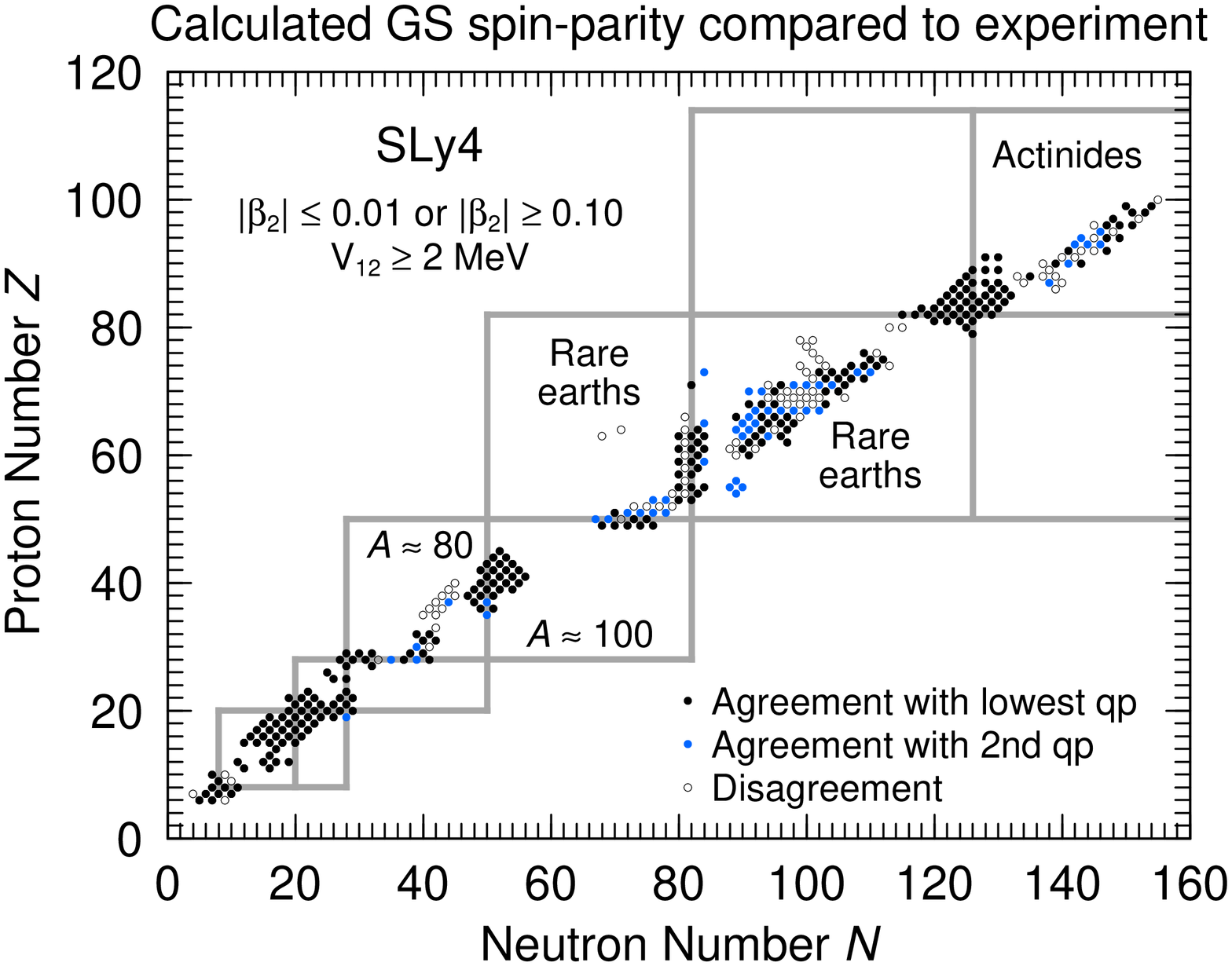}}
\end{center}
\vspace*{-3.5cm}
\caption{(color online). Same as Fig.~\ref{map_agr_SIII_ALL_2} with the SLy4 interaction.
\label{map_agr_SLy4_ALL_2}}
\end{figure*}
\begin{figure*}
\begin{center}
\vspace*{1cm}
\hspace*{-5cm}
\rotatebox{-90}{\includegraphics[width=0.65\textwidth]{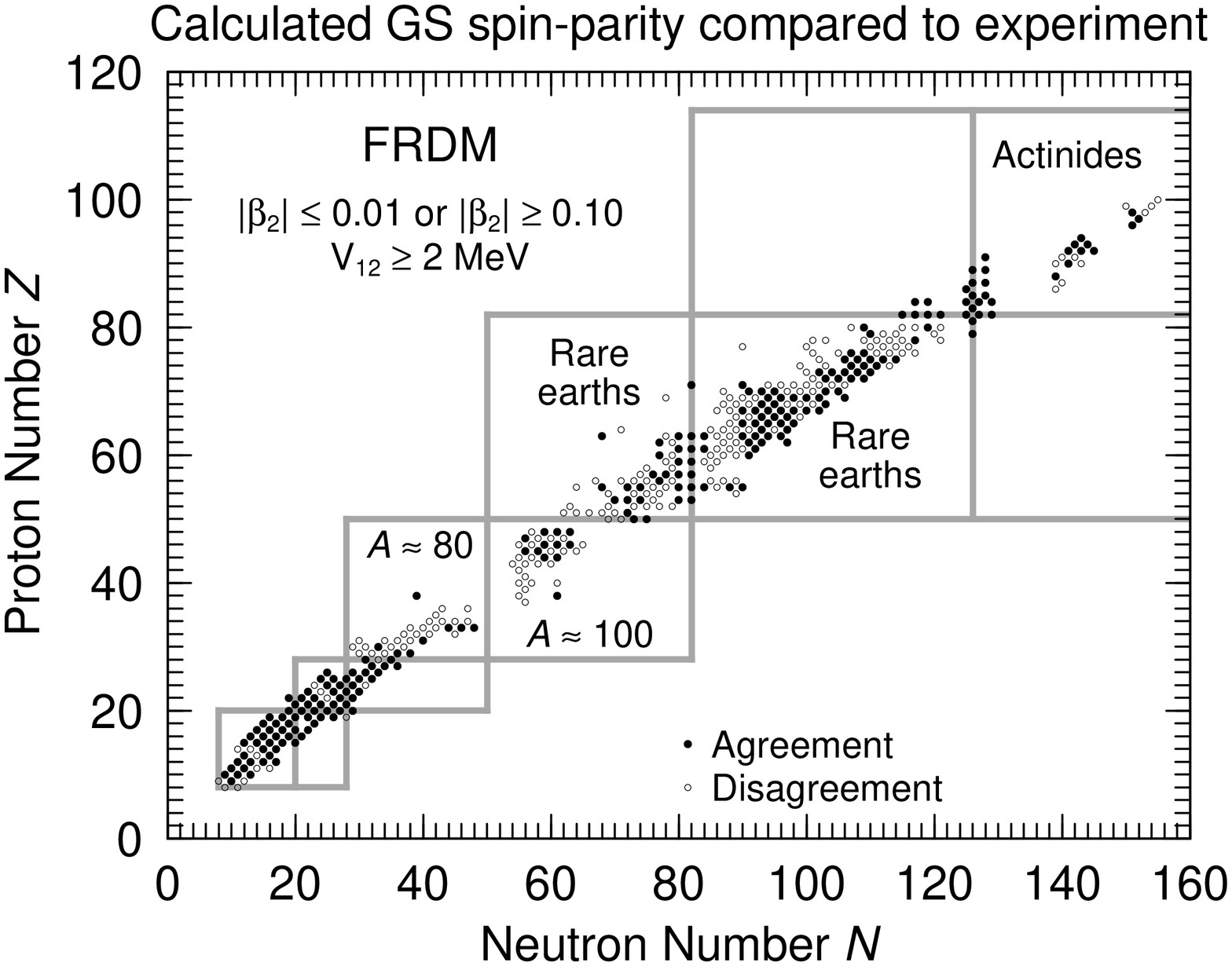}}
\end{center}
\vspace*{-3.5cm}
\caption{Same as Fig.~\ref{map_agr_SIII_ALL_2} within the FRDM model.
\label{map_agr_FRDM_ALL}}
\end{figure*}

Within the HFBCS model, for computation time reasons, we search for the 
lowest-energy solution as a function of $\beta_2$ defined as in 
Ref.~\cite{Moller_mass-table} and calculated exactly for the equivalent 
spheroid having the same quadrupole moment and mean square radius as the 
actual nucleus. The variational character of the HFBCS approach ensures 
that we obtain a minimum with respect to all the other shape degrees of 
freedom (compatible with the restrained symmetries). In principle we 
should search for the lowest local minimum in the whole multi-dimensional 
potential-energy surface, as in the FRDM approach. 
However this is a formidable task in a Hartree--Fock-like approach and not 
necessary in many cases. Indeed for several nuclei in various mass 
regions we have compared HFBCS higher multipole moments from the 
minimization in the $\beta_2$ direction with FRDM ones obtained from 
a minimization in the full deformation space and found similar nuclear shapes.

Since the potential-energy landscape is affected by the pairing 
correlations and since the strength of the seniority force to describe 
these correlations is not known a priori for a given nucleus, we have to 
proceed in an iterative way. Starting with the average set of pairing 
strengths for neutrons and protons, $g_n = 17\mbox{ MeV}/(11 + N)$ and 
$g_p = 17\mbox{ MeV}/(11 + Z)$ respectively, we determine the local 
minima of the HFBCS deformation energy as a function of $\beta_2$. 
Then we calculate the pairing strengths from the solution corresponding 
to the lowest minimum using the pairing model of M\"oller and 
Nix~\cite{Moller_Nix_pairing}. With the new pairing strengths, we 
determine again the lowest local minimum and repeat this procedure until 
the $\beta_2$ value of the GS minimum converges to within 0.01. 
The pairing treatment in the FRDM approach is described in detail 
in Ref.~\cite{Moller_mass-table}.

To solve the Hartree--Fock equations, we diagonalize the one-body 
Hartree--Fock Hamiltonian in the cylindrical harmonic-oscillator basis. 
The truncation of the basis and the approximate optimization of its 
parameters $b$ and $q$ are carried out as in Ref.~\cite{FQKV}. Although 
the oscillator parameter $b$ and the basis size parameter $N_0$ scale as 
$A^{-1/6}$ and $A^{1/3}$, respectively, we choose larger values of 
$N_0$ as a function of $A$, namely values linearly interpolated between 
$N_0 = 12$ for the mass number $A = 10$ and $N_0 = 18$ for $A = 260$, 
and the optimal value of $b$ for the spherical shape is taken to be 
the same for all nuclei, namely $b_0 = 0.475 \mathrm{fm}^{-1}$. We
have checked that the GS radii, quadrupole moments, spins and parities of 
several nuclei across the nuclear chart are not sensitive to the basis 
size when we retain the above value of $b_0$.

We restrict the comparison of the HFBCS results with experimental data 
to nuclei for which the above models are expected to be valid. We therefore 
discard the nuclei for which $0.01 < \beta_2 < 0.1$ (interpreted to be soft) 
and those for which the energy difference $V_{12}$ between the lowest 
two minima is less than 2~MeV (to avoid the possible ambiguities associated with
shape coexistence).

In Table~\ref{tab_res} we show the corresponding percentage of agreement 
and the number of successful spin and parity calculations with the 
SIII, SkM* and SLy4 interactions in the HFBCS model and with the 
FRDM model, for spherical and deformed nuclei. The same type of
information is displayed for each nucleus in the $(N,Z)$ plane,
separately for each model approach (three HFBCS and the FRDM
calculations) on Figs.~\ref{map_agr_SIII_ALL_2}
to~\ref{map_agr_FRDM_ALL}.

In deformed nuclei, single-particle levels often lie very close 
together, which impairs our ability to correctly deduce GS spins 
and parities. This is particularly true in heavy nuclei since 
the single-particle level density increases proportionally to $A$ 
on average. 
Such an uncertainty in the determination of the GS spin and parity 
does not come only from the interactions used, but also from 
a possible reordering of nuclear levels in odd nuclei with respect 
to the order deduced from single-particle ``bare'' spectra caused by 
the coupling of single-particle degrees of freedom with other (collective) 
excitation modes. To estimate this uncertainty in the HFBCS calculations 
we consider not only the lowest-energy quasiparticle but the second lowest 
quasiparticle excitation as well when it lies at most 1~MeV above 
the lowest one. The gray (blue online) dots in Figs.~\ref{map_agr_SIII_ALL_2} 
to \ref{map_agr_SLy4_ALL_2} correspond to the nuclei for which the GS spin 
and parity of the second lowest quasiparticle state agree with the measured 
values, and the number of such cases is indicated in parenthesis in 
Table~\ref{tab_res} for each model and deformation category. In the first 
line for each model in Table~\ref{tab_res} we also indicate in 
parenthesis the percentage of agreement when considering either of 
the two lowest quasiparticle excitations.

As for the coupling with the core rotational mode in deformed nuclei, 
the specific case of single-particle states with $K=1/2$ has drawn our 
attention. Depending on the actual value of the decoupling parameter
$a$, one may have nuclear GS spins different from $1/2$. This happens
when $a<-1$ or $a>4$ and corresponds to about 30\% of the cases where
$K=1/2$. In fact we have calculated $a$ only for the lowest-energy
quasiparticle when it has $K=1/2$, so that the actual spin of the
second lowest quasiparticle when it has $K=1/2$ has not been
calculated. Therefore the number of gray (blue online) dots in
Figs.~\ref{map_agr_SIII_ALL_2} to \ref{map_agr_SLy4_ALL_2} is a lower
limit. However this does not impair the conclusion drawn about the
uncertainty arising from the high single-particle level density in 
deformed nuclei.

As seen from Table~\ref{tab_res} and Figs.~\ref{map_agr_SIII_ALL_2} to 
\ref{map_agr_SLy4_ALL_2} the overall agreement is rather similar for 
all considered Skyrme forces. One may note, however, that the SIII 
parameterization is slightly better than the other two. In all four 
approaches the agreement for the spherical nuclei is much better than 
for deformed nuclei (partly because of the uncertainty mentioned above 
for large single-particle level densities). For the spherical nuclei, 
the FRDM model is more successful than the HFBCS one (90\% as compared 
to 80\%, respectively) but with a much smaller set of nuclei, and yields 
similar results for deformed nuclei, taking only into account the lowest 
quasiparticle state. 

In the mass region $A\leqslant 100$, including all spherical 
and deformed nuclei, the agreement is excellent in all HFBCS cases 
and slightly less good with FRDM (especially for the heaviest nuclei of this 
mass region). In deformed nuclei in or close to the rare-earth region, 
FRDM and SIII calculations are equally good and yield slightly better results 
than the other two Skyrme forces, while in the actinide region the best 
results are obtained from FRDM and SkM* calculations (with a slight advantage 
for the former approach).

To conclude this study, we should stress that given the global 
character of the study (of the order of 400 nuclei involved), the
assumptions made to connect calculated single-particle properties with 
observed spins and parities and the demanding character of such a 
comparison, the agreement can be deemed significant. This gives thus 
a rather good level of confidence in the predictive power of these 
approaches when used in particular contexts. Improvements of this comparative 
study should include a more sophisticated treatment of the coupling of 
single-particle and collective degrees of freedom.

One of the authors (Ph. Q.) acknowledge the Theoretical Division at LANL 
for the excellent working conditions extended to him during numerous visits. 
This work has been supported by the U.S. Department of Energy under 
contract W-7405-ENG-36.


\end{document}